\documentclass[sigconf]{acmart}
\AtBeginDocument{%
  }

\acmYear{2026}\copyrightyear{2026}
\setcopyright{cc}
\setcctype[4.0]{by}
\acmConference[FSE Companion '26]{34th ACM Joint European Software Engineering Conference and Symposium on the Foundations of Software Engineering}{July 5--9, 2026}{Montreal, QC, Canada}
\acmBooktitle{34th ACM Joint European Software Engineering Conference and Symposium on the Foundations of Software Engineering (FSE Companion '26), July 5--9, 2026, Montreal, QC, Canada}
\acmDOI{10.1145/3803437.3806413}
\acmISBN{979-8-4007-2636-1/26/07}

\usepackage{multirow}
\usepackage{booktabs}
\usepackage{tablefootnote}
\usepackage{hyperref}
\usepackage{xcolor}
\usepackage{makecell} % 用于表头换行
\usepackage{xurl}

\begin{document}

%%
%% The "title" command has an optional parameter,
%% allowing the author to define a "short title" to be used in page headers.
\title{Package Dashboard: A Cross-Ecosystem Framework for Dual-Perspective Analysis of Software Packages}

%%
%% The "author" command and its associated commands are used to define
%% the authors and their affiliations.
%% Of note is the shared affiliation of the first two authors, and the
%% "authornote" and "authornotemark" commands
%% used to denote shared contribution to the research.
\author{Ziheng Liu}
% \authornote{Both authors contributed equally to this research.}
\affiliation{%
  \institution{School of Computer Science, Peking University}
  \institution{Key Laboratory of High Confidence Software Technologies, Ministry of Education}
  \city{Beijing}
  \country{China}
}
\email{zihengliu25@stu.pku.edu.cn}
\orcid{0009-0001-2767-6364}

\author{Runzhi He}
\affiliation{%
  \institution{School of Computer Science, Peking University}
    \institution{Key Laboratory of High Confidence Software Technologies, Ministry of Education}
  \city{Beijing}
  \country{China}}
\email{rzhe@pku.edu.cn}
\orcid{0000-0002-6181-6519}

\author{Minghui Zhou}
\authornote{Corresponding author}
\affiliation{%
  \institution{School of Computer Science, Peking University}
  \institution{Key Laboratory of High Confidence Software Technologies, Ministry of Education}
  \city{Beijing}
  \country{China}}
\email{zhmh@pku.edu.cn}
\orcid{0000-0001-6324-3964}

%%
%% By default, the full list of authors will be used in the page
%% headers. Often, this list is too long, and will overlap
%% other information printed in the page headers. This command allows
%% the author to define a more concise list
%% of authors' names for this purpose.
\renewcommand{\shortauthors}{Liu et al.}

%%
%% The abstract is a short summary of the work to be presented in the
%% article.
\begin{abstract}
Recent software supply chain attacks have revealed a critical socio-technical gap in current Software Composition Analysis (SCA) implementations -- isolation among package management ecosystems and open source communities.
This fragmentation poses manual overhead, forcing developers to synthesize scattered data, and potentially undermines the reliability of risk assessments.
To address this, we present Package Dashboard, a cross-ecosystem platform that: 1) synthesizes package metadata, vulnerability feeds, and upstream community health metrics for holistic software supply chain analysis; 
2) provides actionable insights, e.g., recommending resilient alternatives 
to mitigate risky packages.
Through an extensive analysis of 374,000 packages and a user study in three real-world risk mitigation scenarios,
we demonstrate that our approach not only uncovers latent risks but also reduces remediation time by up to 70\%, effectively closing the loop between identifying risks and selecting replacements.

Package Dashboard is publicly available at \url{https://pkgdash.osslab-pku.org},  the code is open source at \url{https://github.com/osslab-pku/PackageDashboard}, and a demonstration video can be found at \url{https://youtu.be/y9ncftP8KPQ}.
\end{abstract}

%%
%% The code below is generated by the tool at http://dl.acm.org/ccs.cfm.
%% Please copy and paste the code instead of the example below.
%%



%%
%% Keywords. The author(s) should pick words that accurately describe
%% the work being presented. Separate the keywords with commas.
%% A "teaser" image appears between the author and affiliation
%% information and the body of the document, and typically spans the
%% page.
% \begin{teaserfigure}
%   \includegraphics[width=\textwidth]{sampleteaser}
%   \caption{Seattle Mariners at Spring Training, 2010.}
%   \Description{Enjoying the baseball game from the third-base
%   seats. Ichiro Suzuki preparing to bat.}
%   \label{fig:teaser}
% \end{teaserfigure}

% \received{20 February 2007}
% \received[revised]{12 March 2009}
% \received[accepted]{5 June 2009}

\maketitle

\section{Introduction}
Open-source software (OSS) has become an integral component of modern software development and enterprise infrastructure. 
% However, the ubiquity is shadowed by the uprising threat of software supply chain attacks. 
% A recent incident in July 2025, where numerous malicious packages were uploaded to the Arch Linux User Repository (AUR) and severely compromised its integrity\cite{Heusel25}, has further amplified industry concerns regarding the fragility of the OSS supply chain.
The exponential growth of open-source software has resulted in a highly complex web of dependencies. This complexity makes the entire ecosystem fragile: a single flaw in a small, widely used package can rapidly cause widespread damage~\cite{npmsec19}. The severity of this fragility was starkly illustrated by the Log4j incident~\cite{Log4jVuln}, which affected nearly 8\% of the Maven ecosystem, and the XZ backdoor~\cite{XZIncident}, which demonstrated the alarming feasibility of Advanced Persistent Threats (APTs) infiltrating core infrastructure through social engineering and long-term dormancy.

% While existing SCA tools offer valuable solutions, they often operate with significant limitations: first, they are typically tailored to a specific ecosystem, making it difficult to apply their insights across different development environments. And the software data is often scattered across various platforms, making it challenging for developers to obtain a comprehensive view of the risks associated with their dependencies.

In the face of the incidents, multiple software composition analysis (SCA) tools have emerged to scan codebases and alert developers to vulnerable dependencies.
Popular examples of these include npm audit~\cite{npmAudit} and GitHub's Dependabot~\cite{dependabot}.
% Software component analysis (SCA) tools have made progress in mitigating software supply chain risks by scanning software's codebase and flagging vulnerable dependencies.
%A wide range of SCA tools exists, but they primarily concentrate on scanning a project’s dependencies to uncover known vulnerabilities. 
Yet, vulnerability detection is only one dimension of ensuring software supply chain resilience. 
Another key dimension is license compliance. Failure to adhere to these open-source licenses can result in  significant legal and ethical consequences. LicenseRec~\cite{licenserec23} addresses this challenge as a knowledge-based open source license recommendation tool that bridges the gap in existing solutions.
% A recent large-scale study of the PyPI ecosystem reveals that this is a prevalent issue~\cite{licenseIncompatibilities}. The complexity of modern software supply chains exacerbates this problem, as 61.3\% of these conflicts originate from transitive dependencies——packages that developers do not directly install but are pulled in by other libraries.
% THE CONSEQUENCES OF LICENSE
% CITE WEIWEI'S TOOL HERE
%$ Yet, the presence of vulnerabilities is not the only factor of supply chain resilience;
% Yet the presence of a vulnerability is only one part of the risk equation; 
Ultimately, community health is the critical indicator of the supply chain's future sustainability. 
An empirical study~\cite{HealthcarePatch22} has proved that the responsiveness and activity strongly correlates with how fast patches are shipped.
%the responsiveness and activity of the upstream community critically determine how quickly patches are released. 
% Reflecting this, the industry has placed growing emphasis on software traceability and community health. 
Reflecting this, tools that analyze a package's community health have emerged, 
including HyperCRX~\cite{HyperCRX25} and Dirty-Waters~\cite{dirtywater25}. 
The former is a GitHub-integrated browser extension that provides real-time visibility into community activities, while the latter explores software provenance by analyzing deprecated tags and inaccessible links.

However, two significant gaps remain in the current tooling landscape: 
\textbf{1) the absence of a unified view}. When performing dependency governance tasks, ranging from selecting an alternative for a deprecated package to choosing a Linux distribution for a cloud infrastructure provider, developers must navigate through multi-source, heterogeneous data to make informed decisions.
Regrettably, there is a disconnect between the mentioned risk dimensions in the current tooling landscape. Even with tool assistance, developers must manually correlate insights from multiple sources, which renders the decision-making process inefficient and error-prone.
\textbf{2) the lack of actionable remediation insights}. 
While existing tools effectively function as ``passive observers'' -- flagging risks like restrictive licenses or project abandonment -- they rarely assist in the discovery of secure alternatives. 
 %% Why this matters in practice
Multiplied by the inconsistent metadata and naming conventions across diverse package management ecosystems,
navigating this fragmented landscape to mine actionable insights and identify 
secure, compliant, and healthy alternatives remains a formidable challenge.
% When a dependency is blocked, developers are often left stranded, forced to manually search for functional equivalents across disparate registries. This process is further complicated by the heterogeneity of software ecosystems (e.g., NPM, PyPI, or RPM-based Linux distributions), where inconsistent metadata and naming conventions make finding a "safe replacement" a time-consuming and error-prone task. 

% The gaps not only complicates XXX
% This fragmentation not only increases the learning curve 
%, but also prevents security teams from establishing a consistent and cross-portfolio risk management strategy.

% To address these challenges,
To bridge the gaps,
we present Package Dashboard, a cross-ecosystem platform 
% designed 
to 
smooth the process from risk identification to remediation.
% bridge the gap between risk identification and remediation. 
To overcome 
% A primary technical challenge is overcoming 
the semantic gap between disparate ecosystems,
our platform systematically normalizes data inputs from five Linux distributions and major language registries (e.g., NPM, PyPI)
into a canonical package and dependency schema.
To present practical remediation insights,
we introduce the Similarity Aggregation Engine to recommend functionally similar, actively maintained and license compliant alternative packages to risky ones.
The engine works by semantic clustering on the topics of GitHub repositories to label ``siblings'' of packages.
% Crucially, beyond visual analytics, we introduce a Similarity Aggregation Engine. By leveraging GitHub Topics and semantic clustering, the system identifies and recommends "sibling" packages—functional alternatives that are actively maintained and license-compliant. 

Package Dashboard is implemented as a dual-component system balancing usability and extensibility.
The Visualization Engine is optimized for low cognitive load, displaying both raw metrics, actionable insights, and mitigation recommendations.
Meanwhile, the Analytics Engine follows a modular, API-first design, ensuring that supply chain insights can be integrated within automated CI/CD workflows.

% The visualization engine is designed with the best effort to reduce cognitive load of developers and DevSecOps engineers, visualizing not only the statistics but also the recommendations.
% Meanwhile, the Analytics Engine follows a modular, API-first design, ensuring that supply chain insights can be integrated within automated CI/CD workflows.

% The back-end follows an API-first design, enabling the integration of these risk insights and recommendations directly into automated pipelines.

% \begin{itemize}  
%   \item We present Package Dashboard, the first tool to provide a unified, cross-ecosystem analysis of open-source software packages, overcoming the fragmentation of existing ecosystem-specific solutions.  
%   \item {\color{red} We implement a novel recommendation mechanism based on GitHub Topics that enables developers to rapidly identify safe, functional alternatives for risky or incompatible packages. } 
%   \item We conduct a large-scale study using Package Dashboard on software packages from five Linux distributions—Anolis, CentOS, Fedora, OpenCloudOS, and OpenEuler—resulting in a dataset of more than 374,000 packages.  
% \end{itemize}

The contributions of this paper are as follows:

\begin{itemize}
  \item We present Package Dashboard, a unified cross-ecosystem platform that: 1) bridges the fragmentation between heterogeneous package management systems and data sources; 2) provides actionable insights, including recommending safe, functional alternative packages to risky ones from topics of GitHub repositories.

  \item To demonstrate the ``bridging information silos'' capability, we integrate over 374,000 packages from five Linux distributions and major package managers (e.g., PyPI, npm), alongside their dependencies, vulnerabilities, licenses, and community health metrics.

  \item To assess the reduction in cognitive load and utility of recommendations, we conduct a user study based on three real-world supply chain risk mitigation tasks. Our results reveal that Package Dashboard reduces the Time-to-Resolution (TTR) by 42\%--70\%, significantly streamlining the decision-making process.

  % \item We implement a novel recommendation mechanism based on GitHub Topics that enables developers to rapidly identify safe, functional alternatives for risky or incompatible packages.
  % \item We conduct a controlled user study demonstrating that Package Dashboard's unified approach and recommendation engine reduce remediation time by up to 70\% compared to traditional disparate workflows, significantly lowering the cognitive load for developers.
  
\end{itemize}

\section{Background}

% Modern software is largely assembled from open-source components, creating a complex supply chain with an opaque attack surface. Ensuring its robustness requires a dual focus: one must address the security risks of the software itself, where 40\% of vulnerabilities resist automated remediation by package managers and patches can take over a year to propagate\cite{npmvuln22}, and also the health of the upstream community, where installing a single NPM package means implicitly trusting 39 maintainers\cite{npmsec19}. This dual perspective is essential for a comprehensive understanding of supply chain risks.

% Beyond these challenges, software ecosystems themselves are highly heterogeneous. Each ecosystem—such as NPM, PyPI, Maven, or Linux distributions—defines its own conventions for package naming, metadata, and dependency resolution. As a result, tools built for one ecosystem often cannot be applied to another, fragmenting the analysis of supply chain security. This diversity complicates both research and practice, as developers must adopt multiple ecosystem-specific tools and security teams cannot establish consistent, cross-portfolio risk assessments.

{\bfseries Software Artifact.} The software artifact is the distributable unit (typically a package) consumed from a package registry (e.g., NPM, PyPI), accompanied by metadata such as name, version, dependencies, and repository links.  Most tools adopt this artifact-centric view,
% , often treating the package as a static object.
often overlooking the evolving risks associated with its upstream community.

\begin{figure}[h] \centering \includegraphics[width=0.7\linewidth]{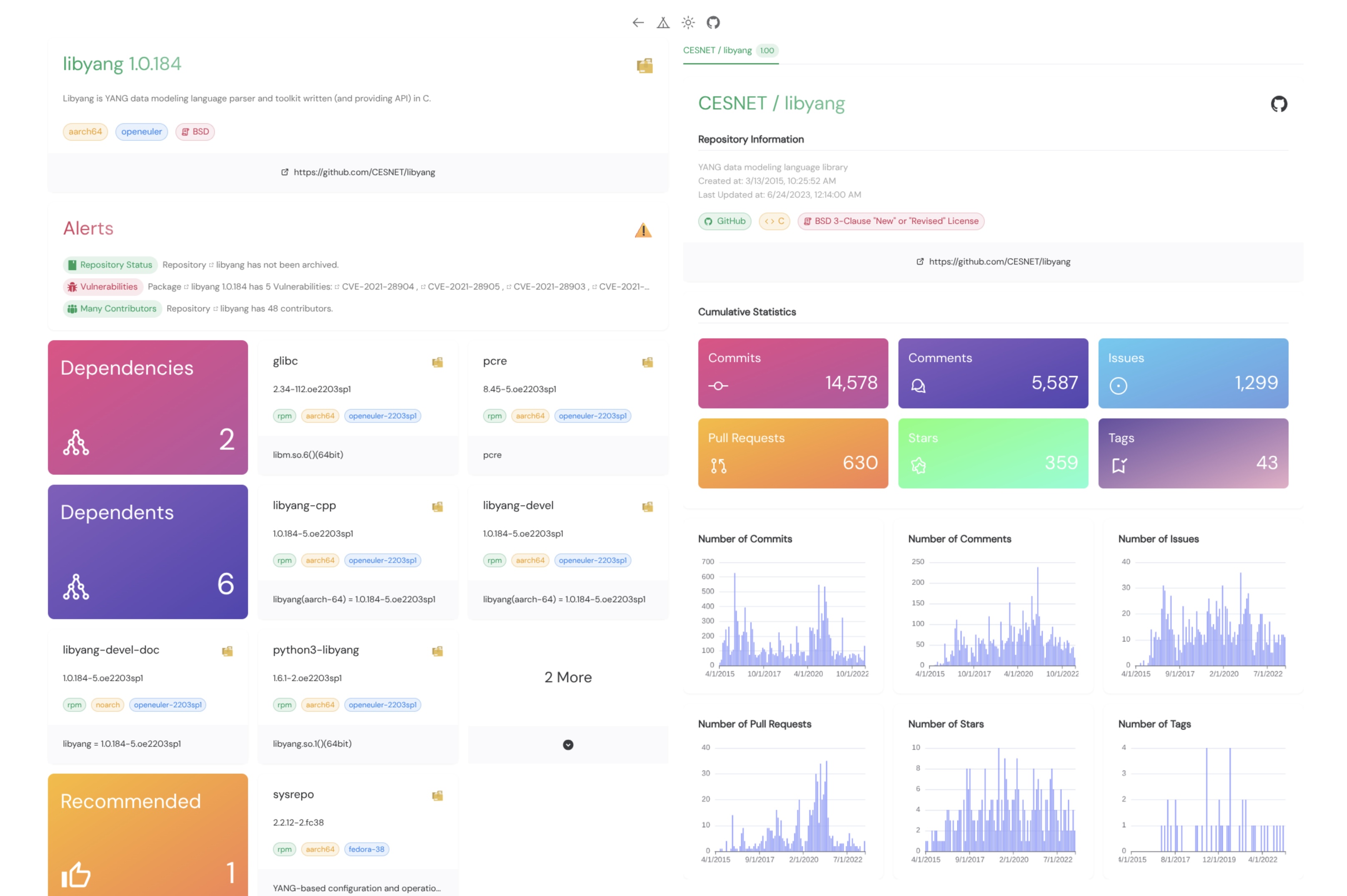} \caption{Package Dashboard Visualizing the libyang Package} \label{package_dashboard} \end{figure}

\begin{figure}[h] \centering \includegraphics[width=0.6\linewidth]{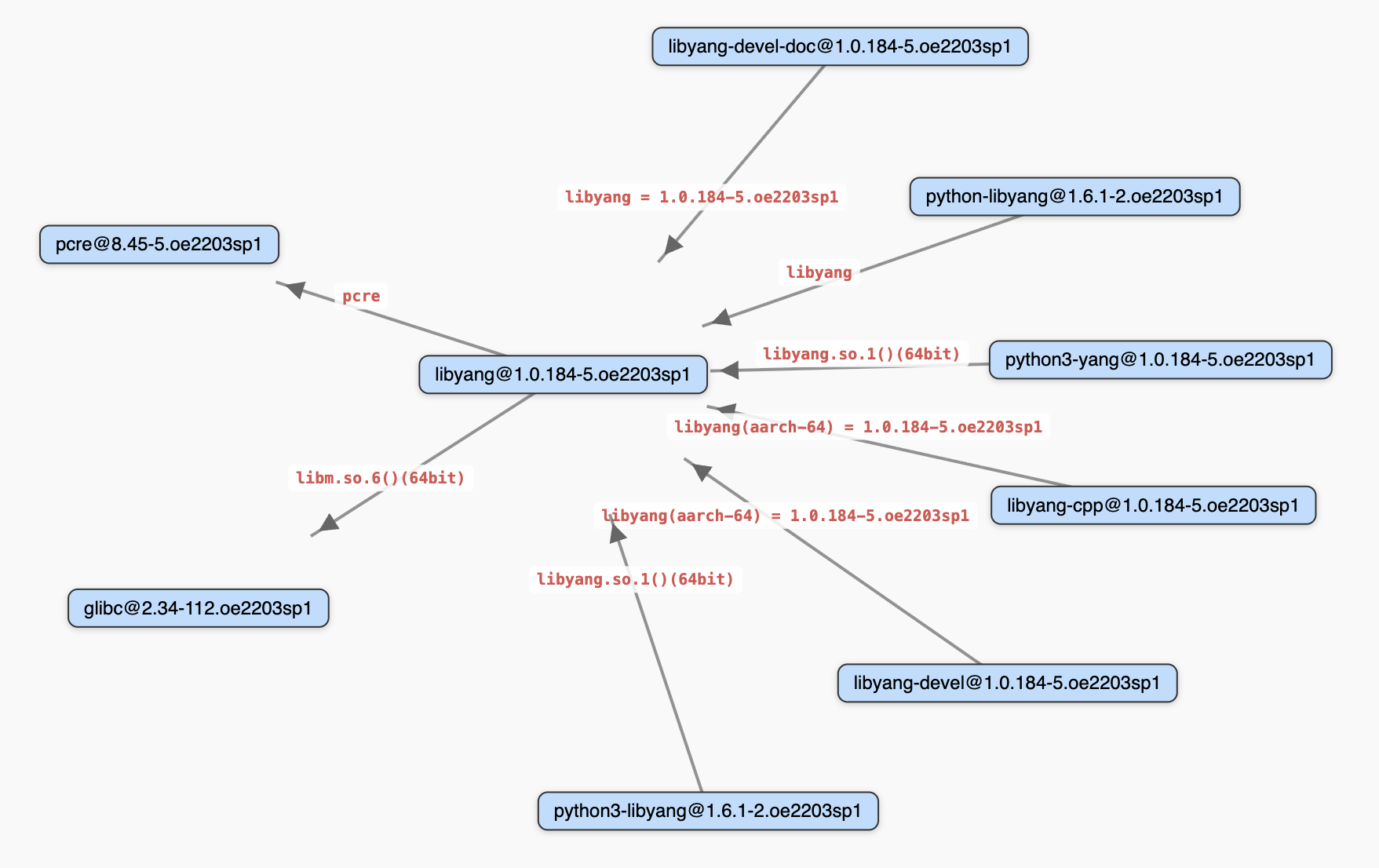} \caption{The Dependency Graph of libyang} \label{deps} \end{figure}

{\bfseries Upstream Community.} The upstream community—developers, repositories, and workflows behind an artifact—is its birthplace and nurturing environment. Community health, from responsiveness to maintenance practices, directly affects the artifact's security and viability. Yet most tools analyze artifacts in isolation, overlooking the ecosystem that sustains them. 
% For example, the success of automated dependency updates often depends on community response, as developers are receptive to bot-generated pull requests like Dependabot\cite{dependabot}. 
This highlights the need for platforms that integrate both artifact and community perspectives into a unified view of supply chain health.

{\bfseries The Gap.} The primary challenges in software composition analysis spanning over multiple ecosystems
stem from the heterogeneity and inconsistency of the data.
Each ecosystem creates unique package formats
and metadata structures, and
employs unique naming schemes.
%and metadata structures. 
%stem from inconsistent package formats and data heterogeneity, as each ecosystem employs unique naming schemes and metadata structures. 
While current efforts like the Package URL (PURL) specification~\cite{PkgUrl} partially standardizes the identification of the packages, 
the gap between software artifacts and their upstream communities across diverse ecosystems remains unfilled.
% a unified platform that holistically analyzes both software artifacts and their upstream communities across these diverse ecosystems is notably absent in the current literature.

% This fragmentation highlights the primary challenges in multi-ecosystem Software Composition Analysis: inconsistent package formats and data heterogeneity. Different ecosystems utilize unique naming schemes and metadata structures, making it difficult to track and analyze software components uniformly across platforms. To address inconsistent identification, the Package URL (PURL) specification was introduced as a standard for universally identifying software packages. Similarly, creating a common structured data representation is a necessary step to handle the heterogeneity of information from diverse sources. These foundational challenges must be overcome to enable any meaningful cross-ecosystem risk assessment.

% Additionally, we find that a unified platform for the holistic analysis of cross-ecosystem software artifacts and their communities is notably absent in the current literature.

\begin{figure*}[h]
  \centering
  \includegraphics[width=0.65\linewidth]{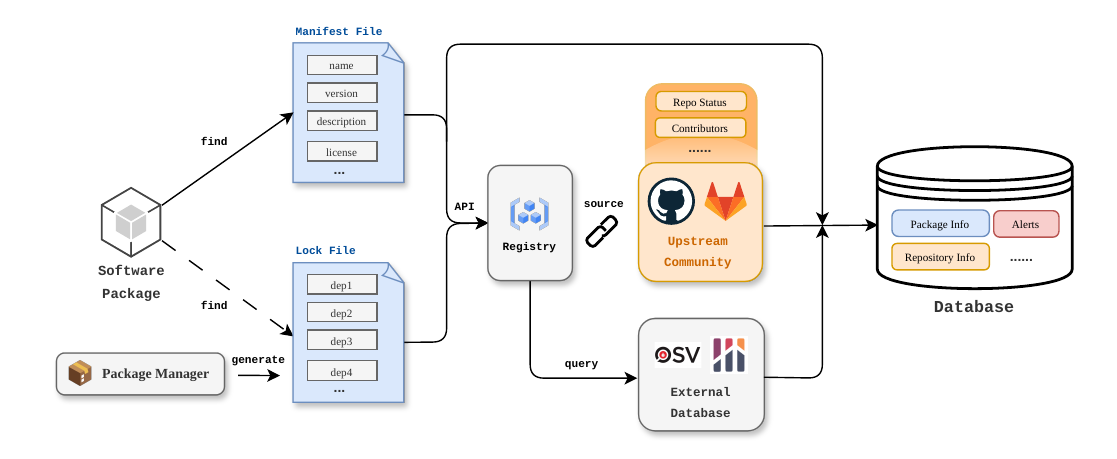}
  \caption{The Workflow of the Analytics Engine}
  \label{analyzer}
\end{figure*}

\section{Implementation}

% Package Dashboard is implemented with a decoupled front-end and back-end architecture. The back-end is responsible for the collection, analysis, and persistence of data , while the front-end receives and visualizes these data to help developers understand the risks within the software supply chain from multiple perspectives.

\subsection{Visualization Engine}

The Visualization Engine of Package Dashboard is designed with three goals in mind: 1) to present a dual-perspective view that integrates both software artifact and community health data; 2) to provide a consistent user experience across different ecosystems. 3) given that the underlying data is inherently multifarious and heterogeneous, 
% the fundamental design goal 
the key 
is to reduce the user's cognitive load by transforming complex information into clear, actionable insights. 
% To achieve this, we established a core design platform centered on three principles:
To achieve this, we implemented three design strategies:

{\bfseries Unified Dual-Perspective Layout.} The cornerstone of our UI is a consistent two-panel layout that directly mirrors our dual-perspective analysis model. 
The left-hand panel (Figure~\ref{package_dashboard}) presents software artifact details essential to DevSecOps engineers: dependencies, vulnerabilities, and licenses.
% As shown in Figure~\ref{package_dashboard}, the 
% % left-hand side 
% the left
% is 
% % always 
% dedicated to the Software Artifact, displaying concrete details like dependencies, CVEs, and license information.
The right-hand upstream community panel visualizes historical community health metrics, including repository activity and contributor statistics.
% The right-hand side consistently represents the Upstream Community, presenting dynamic health metrics such as repository activity and contributor statistics.
This strict modularization allows users to intuitively grasp a package's complete risk profile at a glance, regardless of its ecosystem of origin.

{\bfseries Interactive Relationship Visualization.} 
Unraveling a complex dependency network can be a major challenge in supply chain analysis. 
To make these relationships clear, we render an interactive dependency graph using {\texttt{D3.js}}, as shown in Figure \ref{deps}. 
% This is not merely a static image; 
Users can inspect the statistics of the dependencies by simply clicking any node on the graph.
% Users can click on any node to seamlessly navigate to that component's own dashboard page. 
% This feature empowers users to conduct in-depth, exploratory analysis of the entire supply chain 
% from a single starting point.
% with minimal overhead of navigation.

{\bfseries Contextualized Risk Highlighting.} The UI automatically highlights and contextualizes diverse risks, moving beyond simple issue lists. It presents conventional threats like CVEs alongside often-overlooked community warnings, such as an ``archived'' repository status, within their appropriate panels.
% This approach enables users to make more informed and nuanced decisions.

\subsection{Analytics Engine}

The primary challenge in cross-ecosystem analysis is overcoming the immense heterogeneity of toolchains and data schemas. The Analytics Engine of Package Dashboard is engineered as 
%a sophisticated orchestration layer 
an abstract layer
to tackle this complexity. It establishes a robust, abstract workflow that normalizes disparate inputs and exposes a single, consistent API, effectively creating a lingua franca for cross-ecosystem security analysis.

{\bfseries Ingestion and Dependency Resolution.} 
This module provides a universal entry point for dependency resolution.
The system extracts the dependency tree from both manifest and lock files. In the absence of a lock file, it performs a simulated clean installation within an isolated virtual environment. This strategy ensures the construction of a precise and complete transitive dependency tree under all circumstances.
% This module provides a common entrypoint for dependency resolution by integrating with the native package manager of any given ecosystem. 
% It interprets package metadata from a Manifest File and a Lock File. If a Lock File is absent, a clean installation is simulated within a virtual environment to generate one, ensuring a precise, transitive dependency tree is always constructed.

{\bfseries Metadata and Upstream Community Analysis.} The system consolidates information by communicating with various package registries and source code platforms. It parses the manifest file to extract a consistent set of metadata and 
repository links.
%community health metrics. 
Besides, the component 
% acts as a centralized conduit for data enrichment from third-party services. 
connects a number of supply-chain related data sources for a more credible and comprehensive insight.
The component queries the librarie.io~\cite{katz2018libraries} dataset for metadata completion and cross-validation.
Vulnerabilities are obtained from OSV~\cite{GoogleOSV}.
It analyzes license compliance based on the work of LicenseRec ~\cite{licenserec23}.

% It queries cross-ecosystem databases like OSV for known vulnerabilities and Libraries.io for supplementary package information, with cross-validation to ensure data integrity. The license compatibility analysis is derived from LicenseRec ~\cite{licenserec23}.

{{\bfseries Similarity Aggregation Engine.} 
To address the challenge of high switching cost
when dependencies become abandoned or license-incompatible, the component implements a recommendation logic. 
By harvesting GitHub Topics and repository descriptions via the upstream community API, the system constructs a semantic cluster of packages. For a given artifact, the engine queries the database for ``sibling'' packages that share a significant overlap of functional topics.}

{\bfseries Data Structuring and Persistence.} Finally, all gathered intelligence is harmonized into a single data model and persisted in a central database. This procedure organizes diverse information into consistent entities -- Package Info, Repository Info, and Security Alerts -- which creates an source of truth for the API. 
Table \ref{tab:info} summarizes the information provided by Package Dashboard, with their sources and purposes.

\begin{table*}[]
  \caption{Summary of Information Types, Sources, and Purposes}
  \label{tab:info}
  \begin{tabular}{lll}
    \toprule
    \bfseries Type                                                                                                                                         & \bfseries Source                                  & \bfseries Purpose                \\
    \midrule
    Package Name, Version, Description                                                                                                                     & Manifest File                                     & \multirow{5}{*}{Transparency}    \\
    License Information                                                                                                                                    & Manifest File                                     &                                  \\
    Dependency (Direct \& Transitive)                                                                                                                      & Lock File / Dependency Resolver                   &                                  \\
    Dependents (Reverse Dependencies)                                                                                                                      & Package Dashboard Internal Analysis             &                                                                                      \\
    Community Activity Metrics (Commits, Issues, PRs, etc.)                                                                                                & Upstream Community Repository                     &                                  \\
    \midrule
    Community Activity Historical Trends                                                                                                                   & Upstream Community Repository                     & \multirow{4}{*}{Trustworthiness} \\
    Repository Status (e.g., Archived)                                                                                                                     & Upstream Community Repository                     &                                  \\
    Known Vulnerabilities (CVEs)                                                                                                                           & OSV Database &                                  \\
    License Compatibility Alerts                                                                                                                           & LicenseRec &                                                                                      \\
    \midrule
    Source Code Repository URL                                                                                                                             & Package Registry                                  & Traceability                     \\
    \bottomrule
  \end{tabular}
\end{table*}

\begin{table*}[htbp]
  \centering
  \caption{User Study Results: Efficiency and Satisfaction Evaluation}
  \label{tab:user_study_results}
  \begin{tabular}{lccc}
    \toprule
    \textbf{Metric} & \makecell[c]{\textbf{Scenario A} \\ (License Conflict)} & \makecell[c]{\textbf{Scenario B} \\ (Unpatched CVE)} & \makecell[c]{\textbf{Scenario C} \\ (Distro Selection)} \\
    \midrule
    Baseline Avg. Time       & 7.4 mins         & 9.8 mins         & 10.6 mins \\
    Dashboard Avg. Time      & \textbf{4.3 mins} & \textbf{3.5 mins} & \textbf{3.2 mins} \\
    Efficiency Improvement   & 42\%        & 64\%        & 70\% \\
    User Satisfaction Score    & 4.8 / 5.0         & 4.7 / 5.0         & 4.9 / 5.0 \\
    \bottomrule
  \end{tabular}
\end{table*}

{\section{User Study}}

To demonstrate the effectiveness of Package Dashboard in facilitating rapid remediation, we conducted a controlled user study with 12 participants, all possessing a minimum of two years of relevant experience. The study measured Time-to-Resolution (TTR) across three representative supply chain risk mitigation tasks: 1) identifying a license-compliant alternative for a dependency blocked by strict corporate policy; 2) finding a maintained substitute for an abandoned component with unpatched critical vulnerabilities; 3) selecting the optimal Linux distribution by comparing security patch responsiveness across five different ecosystems.

In the control group, participants relied on traditional workflows, manually correlating data from disparate registries, source repositories, and vendor errata pages. This process proved particularly labor-intensive for the distribution selection task, where verifying backported patches required navigating inconsistent vendor advisories. In contrast, the experimental group utilized Package Dashboard’s Similarity Aggregation Engine to instantly cluster and rank functional alternatives by community health and license compatibility. Similarly, the unified cross-ecosystem search allowed participants to directly compare the maintenance status of core libraries across distributions without leaving the platform.

To 
% capture the qualitative 
quantify  
% impact of the tool, 
the user experience of Package Dashboard,
we supplemented the time-based metrics with a post-task user experience survey. Upon completing each scenario, participants were asked to rate their satisfaction on a 5-point Likert scale (1 = Strongly Dissatisfied, 5 = Strongly Satisfied), specifically evaluating the clarity of the provided information and their confidence in the remediation decision. The User Satisfaction Score presented in Table \ref{tab:user_study_results} represents the mean value of these ratings for each scenario, serving as a proxy for the perceived reduction in cognitive load and the trustworthiness of the system's recommendations.

As shown in Table \ref{tab:user_study_results}, Package Dashboard significantly reduced remediation time across all scenarios. The most dramatic improvement (70\%) was observed in the distribution selection task, validating the utility of our normalized multi-ecosystem dataset. This substantial performance gap stems from the elimination of manual context switching; whereas baseline participants were forced to navigate and interpret inconsistent advisories from divergent vendor portals, the Dashboard presented an instantly harmonized view. Furthermore, the efficiency gains in alternative package selection tasks indicate a fundamental shift in the remediation workflow: the Similarity Aggregation Engine transforms the process from an open-ended external search into a streamlined internal verification task, significantly lowering the cognitive threshold for decision-making.

\section{Limitation}

First, Package Dashboard trusts the metadata extracted from dependency manifests without exhaustive verification. Consequently, the credibility of analysis results relies on the credibility of the manifests, which is proven susceptible to human errors, such as inheriting template defaults or declaring invalid values.
~\cite{10.1145/3660822}.
% First, our platform relies on metadata extracted from package manifests without exhaustive verification. Consequently, data quality is susceptible to developer errors, such as the use of default values or invalid entries, which may propagate into the analysis results.
Second, our remediation recommendation relies on manually labeled GitHub topics for semantic clustering.
Theoretically this method is less precise than a comprehensive textual analysis utilizing text embeddings or large language models.
This decision is a trade-off to scale up the dataset to 374,000 packages, while keeping the cost and computational latency acceptable.
The user study validates that our lightweight approach does not hurt the user experience.

% Second, our Similarity Aggregation Engine utilizes GitHub Topics for clustering. 
% While this method is less precise than state-of-the-art Large Language Model (LLM) based recommendation algorithms or semantic text similarity, it was selected to ensure system scalability.
% Applying LLM-based analysis to a dataset of 374,000 packages would present unsustainable computational costs and latency, making the current heuristic a necessary trade-off for large-scale processing.
\vspace{-.2in}
\section{Conclusion}

In this paper, we presented Package Dashboard, a comprehensive cross-ecosystem supply chain analysis platform.
Package Dashboard bridges the critical gap between software artifact and upstream community. 
We demonstrate its generalizability by integrating over  374,000 packages from 
five Linux distributions and major programming language registries.
Unlike traditional SCA tools that often require more manual effort after flagging issues,
Package Dashboard yields practical remediation insights.
The platform utilizes GitHub Topics to proactively recommend secure, license-compliant alternatives for risky dependencies. 
% Supported by a comprehensive dataset of over 374,000 packages across five Linux distributions. 
The user evaluation demonstrates that this integration 
accelerates decision-making and drives a fundamental shift in remediation workflows.
By eliminating manual context switching and streamlining the discovery of substitutes, Package Dashboard reduces decision-making time by up to 70\% and significantly lowers the cognitive load for developers. Ultimately, by contextualizing technical risks with community insights and providing actionable remediation paths, Package Dashboard empowers DevSecOps teams to move beyond passive observation to active, evidence-based risk mitigation, fostering a more transparent and resilient open-source ecosystem.

\begin{acks}
This work is supported by the Fundamental and Interdisciplinary Disciplines Breakthrough Plan of the Ministry of Education of China (No.\ JYB2025XDXM101) and the National Natural Science Foundation of China (No.\ 62332001).

\end{acks}

%%
%% The next two lines define the bibliography style to be used, and
%% the bibliography file.
\bibliographystyle{ACM-Reference-Format}
\bibliography{pd-ref}

\end{document}